\documentclass[fleqn,twoside]{article}
\usepackage{espcrc2}

\usepackage{graphicx} 

\usepackage[figuresright]{rotating}

\newcommand{\AmS}{{\protect\the\textfont2
  A\kern-.1667em\lower.5ex\hbox{M}\kern-.125emS}}

\title{Radiation Hardness and Linearity Studies of CVD Diamonds}

\author{
T.Behnke\address[DESY]{Deutsches Elektronen Synchrotron,
        Notkestra\ss e 85, 22607 Hamburg, Deutschland}, 
M. Doucet\addressmark[DESY]\address[NIST]{ now at
	NIST Center for Neutron Research,
	100 Bureau Drive, Stop 8562,
	Gaithersburg MD 20899-8562, USA.}, 
N. Ghodbane\addressmark[DESY]\thanks{Speaker at the 8$^{th}$ Topical Seminar on Innovative
 Particle and Radiation Detectors Siena, 21-24 October 2002
},
A. Imhof\addressmark[DESY]
}

\begin{document}

\begin{abstract}
\vspace{1pc}
We report on the behavior of CVD diamonds under intense electromagnetic radiation and on the response
of the detector to high density of deposited energy. Diamonds have been found to remain unaffected after
 doses of 10 MGy of MeV-range photons and  the diamond response to energy depositions of
 up to 250 GeV / cm$^3$ has been found to be linear to better than 2 \%. These observations
 make diamond an attractive detector material for a calorimeter in the very forward
 region of the detector proposed for TESLA.
\end{abstract}
\maketitle

\section{Introduction}

With the establishment of the Chemical Vapor Deposition (CVD) growth
processes, diamond detectors started to be extensively investigated
for their use for particle detection at future high energy and nuclear
physics experiments. The developments were driven by the need for a
radiation detector, particularly at the upcoming generation of
experiments at hadron colliders like the Large Hadron Collider
(LHC). The main problem diamond is faced with is its small charge
collection efficiency, resulting in bad signal to noise ratios. Most
of the studies, focused on the hadronic radiation hardness properties,
have shown that diamond detectors suffer some radiation damage for
fluences of neutrons, protons or pions above approximately
$10^{15}$/cm$^2$~\cite{theses}.\\
At a possible future high energy
electron positron collider like TESLA~\cite{tesla:tdr}, the detectors
in the very forward region are subjected to very high radiation doses,
mostly electromagnetic radiation produced as beamstrahlung. Doses as
high as several MGy per year are expected. Recently a number of
studies have been published on the use of diamond in these regions as
a possible detector material~\cite{behnke:radhard}. \\
In this paper, after a
description of the principle of operation of diamonds and the
definition of some important parameters, we summarize the results
concerning the radiation hardness, the Thermally Stimulated Current
measurements (TSC) and the linearity tests.

\section{Principle of Operation}
\label{section:principle}
Diamond has a number of properties which make it an attractive
material for use in particle detection applications. The ionization
produced by energetic charged particles passing through a thin diamond
film, typically of the order of 300-500~$\mu$m thick, creates about 36
electron-hole pairs per $\mu m$ of diamond. By applying a potential
difference (typically 1V/$\mu m$) between the two electric contacts
the charges start to migrate toward them. Due to impurities and
dislocations in the diamond, some of the migrating charges are trapped
and may contribute to space charge, which builds up and polarizes the
diamond crystal. The charge induced on the contacts $Q_{{\rm
induced}}$ is then smaller than the total charge created in the
diamond $Q_{{\rm deposit}}$. \\ A widely used figure of merit of the
diamond material is its charge collection efficiency defined as
$\epsilon = Q_{{\rm induced}} / Q_{{\rm deposit}} $. For current
diamond material the charge collection efficiency is typically below
$50\%$.\\
If the diamond is exposed to high levels of ionizing radiation,
additional defect may be created in the diamond lattice, which may
trap charge. These defects may be caused by impurities in
the material (e.g. Nitrogen) or by structural damage to the lattice
(e.g. dislocations, etc.). Several methods exist to study the defects
and to better understand the mechanism of charge trapping in diamond. A
particular powerful method one is the method of Thermally Stimulated
Current (TSC). In this method, after having filled the traps at room
temperature, the diamond is heated up. The rise of the sample
temperature induces thermal detraping at a rate depending on the
temperature and on the energy levels of the traps. A current
proportional to the trap density and to the release rate is then
observed between the contacts of the diamond sample. The TSC
dependence on the temperature gives information about the energy
levels and the density of the impurities in the diamond. \\ 
In
addition to maintaining a good charge collection efficiency, a
detector to be used in a high radiation rate environment also has to
show a linear response to large energy deposition. This is
particularly important for the application of this material in a
possible forward detector at a linear collider, where energy deposits
over many orders of magnitude are to be expected.\\
In the following sections, we summarize studies and present results
for both radiation and linearity issues for diamond detectors. For a
more detailed description of the analysis see reference~\cite{behnke:radhard}.

\section{Radiation Hardness}
\label{section:hardness}

\begin{figure}[h!tb]
\includegraphics[width=7.5cm]{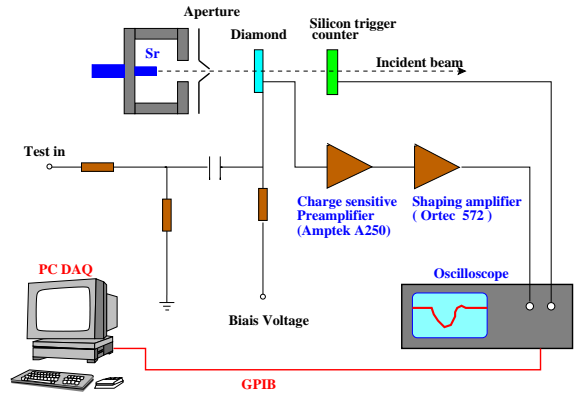}
\caption{Charge Collection Efficiency Measurement Setup.}\label{ccd-setup.eps}
\end{figure}

The response of CVD diamonds to uniform ionization densities has been
measured using the setup shown in figure~\ref{ccd-setup.eps}. The
diamond samples were metalized with Ti/Pt/Au electrodes on both sides
and placed in front of a collimated $^{90}$Sr source which delivers
electrons up to a maximum energy of 2.28 MeV.  
In order to guarantee
that only signals from $^{90}$Sr decay electrons which penetrate the
diamond are recorded, a silicon trigger counter is placed behind the
diamond sensor. The diamond signal is amplified using a charge
integration amplifier circuit and can be viewed on a digital scope or
recorded on a computer for future analysis. From the measured charge,
one can estimate the charge collection efficiency.

\noindent Three Diamond samples were exposed to electromagnetic radiation
with two different ranges: below and above the threshold for non ionizing
damages. A 10 KeV photon beam provided by the Hasylab facility
at DESY with an average dose rate of 14 Gy/s was used to
irradiate two diamond samples for doses up to 1.6 and 6.8 MGy. A third
diamond sample was sent to a $^{60}$Co irradiation facility which
provides photons with energies of 1.17 and 1.33 MeV. The diamond
sample was irradiated with a dose of 10 MGy.
\begin{figure}[h!tb]
\includegraphics[width=7.5cm]{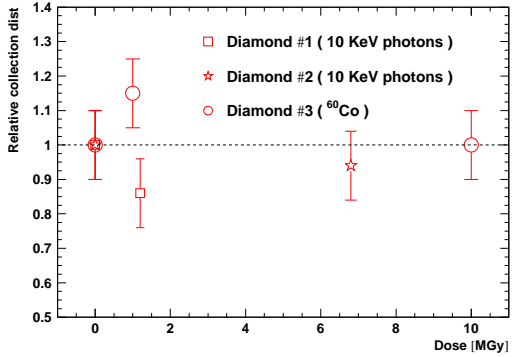}
\caption{Ratio of the collection efficiencies measured after irradiation to the collection distance
 measured before irradiation for several diamonds and irradiation periods.}\label{colldose_rel.eps}
\end{figure}

\noindent Figure \ref{colldose_rel.eps} show the total relative charge collection  efficiency for the three diamond samples for different irradiation doses. No  indication of degradation of the diamond quality as a function of the radiation dose has been observed.

\section{Thermally Stimulated Current}
\label{section:tsc}
\begin{figure}[h!tb]
\includegraphics[width=7.5cm]{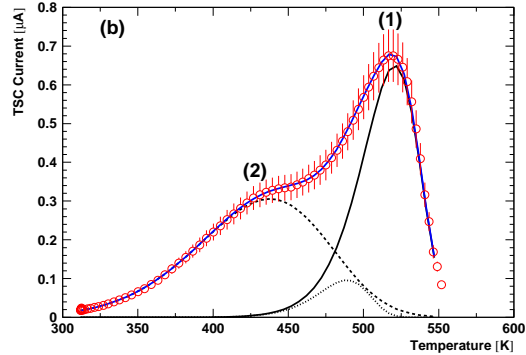}
\caption{TSC spectra and associated deconvolution analysis. The deconvolution yields two energy levels at 1.19 eV (peak 1) and 0.35 eV (peak 2).}\label{r46e23.eps}
\end{figure}

Using the TSC method several diamond samples have been studied for
their defect structure. After irradiation the samples were exposed to photons from a weak 10 KeV source to create electron-hole pairs in the sample.\\ 
To measure the TSC, a nominal voltage of 50 V
was applied on one contact of the diamond during measurement and
irradiation periods.  The other contact was connected to a Keithley
6514 nano-A meter to measure the current. To generate the TSC, a
remote-controlled heating element was used.  The temperature was
monitored using a thermocouple element read by a voltmeter. \\
Before heating the diamond, a fixed period of irradiation was used to
create electron-hole pairs in the diamond and fill traps. For this
purpose, the 10 KeV photon beam was directed on a 100 $\times$ 100
$\mu$m$^2$ slit. A remote-controlled shutter was placed between the
detector and the slit to switch the beam on and off between data
taking periods. For each acquisition sequence, an irradiation period
of 60 s, well below the time period over which saturation effects
become important, was done before the TSC curve was recorded. To study
the influence of different areas on the diamond on the measurement an
area of 1~mm$^2$ was scanned in steps of 100 $\times$ 100
$\mu$m$^2$.\\
 Figure \ref{r46e23.eps} shows a sample TSC curve of the
evolution of the measured current with temperature. A fit to the TSC
curves was performed to evaluate the energy levels, the frequency
factors and the total density of traps. The deconvolution analysis
yields to two energy levels equal to 0.35 and 1.19 eV.

\section{Response Linearity}
\label{section:linearity}

\begin{figure}[h!tb]
\includegraphics[width=7.5cm]{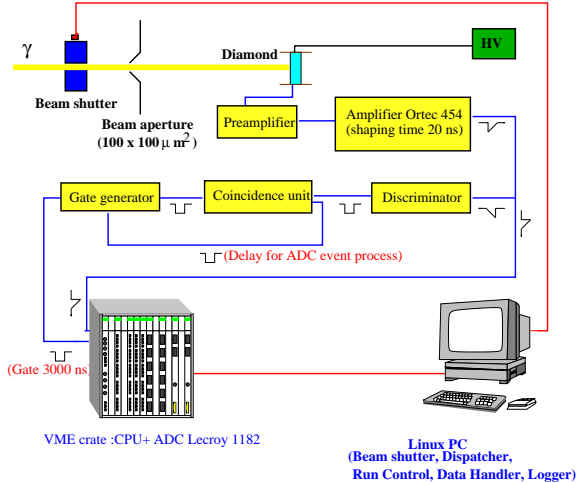}
\caption{Linearity Setup}\label{linearity.eps}
\end{figure}

The linearity of the response of the diamond detector was tested using 
the Hasylab synchrotron radiation facility at DESY.  A diamond detector 
was placed in a 10~KeV photon beam  which was intense enough to provide a high energy density 
two orders of magnitude larger than what is expected at the TESLA Luminometer. 
The beam intensity was about 2000 photons per bunch on a 
$300\times 300$~$\mu$m$^2$ area.  The beam was shone on a diamond
surface of about 1~mm$^2$, defined by a copper mask.
The photon flux on the detector for the maximum beam
current (about 80~mA) corresponds to a deposited energy
density in the diamond of about 7.5~GeV/cm$^2$ for a 300~$\mu$m 
thick diamond.\\
Figure \ref{linearity.eps} shows a sketch of the experimental setup used for the linearity studies. 
The diamond detector was read out using an 
Amptek~A250
pre-amplifier, that was 
followed by an Ortec amplifier/shaper with a 50~ns shaping time.  
The output signal of the amplifier/shaper was sent to a Lecroy~1182 ADC.

\begin{figure}[htb]
\includegraphics[width=7.5cm]{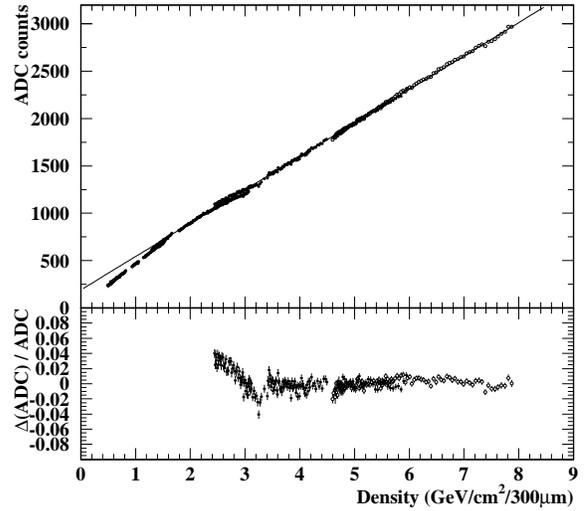}
\caption{
     Diamond signal (ADC counts) as a function of the deposited 
     energy density.
     The line shows the fit result.
     The bottom plot shows the relative difference $\Delta$(ADC)/ADC 
     between the measurement and the linear fit.}\label{linear_dens_del.eps}
\end{figure}

\noindent Figure~\ref{linear_dens_del.eps} shows the variation of the measured
signal as a function of the energy density in the diamond. One can clearly notice the linear relationship between the energy deposited in the diamond detector and the total collected charge. For low energy densities, one sees a non-linear behavior,  ascribed to the pedestal. The bottom plot on figure~\ref{linear_dens_del.eps} shows the 
relative difference $\Delta$(ADC~counts)/(ADC~counts) between the data 
and the linear fit to these data.  \\
These measurements show that
the diamond detector is linear to better than 2\% up to 7.5~GeV/cm$^2$ 
for a detector thickness of 300~$\mu$m.

\section{Conclusions}
Properties of CVD diamond detectors like their radiation hardness, their characterization using the TSC method and their linearity response to large amount of electromagnetic deposited energy, have been studied. Diamonds detectors have been found to exhibit a linear response to better than 2\% for energy depositions of up to 250 GeV / cm$^3$.

\end{document}